\PassOptionsToPackage{unicode}{hyperref}
\PassOptionsToPackage{hyphens}{url}
\PassOptionsToPackage{dvipsnames,svgnames,x11names}{xcolor}

\documentclass[
  12pt]{article}

\usepackage{amsmath,amssymb}
\usepackage{algpseudocode}
\usepackage{authblk}
\usepackage{algorithm}
\usepackage{subcaption}
\usepackage{mathrsfs}
\usepackage{comment}
\usepackage[scr=rsfs, cal = boondox]{mathalfa}
\usepackage{iftex}
\usepackage{url}
\usepackage{doi}
\ifPDFTeX
  \usepackage[T1]{fontenc}
  \usepackage[utf8]{inputenc}
  \usepackage{textcomp}
\else
  \defaultfontfeatures{Scale=MatchLowercase}
  \defaultfontfeatures[\rmfamily]{Ligatures=TeX,Scale=1}
\fi
\usepackage{lmodern}
\ifPDFTeX\else  
\fi
\IfFileExists{upquote.sty}{\usepackage{upquote}}{}
\IfFileExists{microtype.sty}{
  \usepackage[]{microtype}
  \UseMicrotypeSet[protrusion]{basicmath}
}{}
\makeatletter
\@ifundefined{KOMAClassName}{
  \IfFileExists{parskip.sty}{
    \usepackage{parskip}
  }{
    \setlength{\parindent}{0pt}
    \setlength{\parskip}{6pt plus 2pt minus 1pt}}
}{
  \KOMAoptions{parskip=half}}
\makeatother
\usepackage{xcolor}
\makeatletter
\ifx\paragraph\undefined\else
  \let\oldparagraph\paragraph
  \renewcommand{\paragraph}{
    \@ifstar
      \xxxParagraphStar
      \xxxParagraphNoStar
  }
  \newcommand{\xxxParagraphStar}[1]{\oldparagraph*{#1}\mbox{}}
  \newcommand{\xxxParagraphNoStar}[1]{\oldparagraph{#1}\mbox{}}
\fi
\ifx\subparagraph\undefined\else
  \let\oldsubparagraph\subparagraph
  \renewcommand{\subparagraph}{
    \@ifstar
      \xxxSubParagraphStar
      \xxxSubParagraphNoStar
  }
  \newcommand{\xxxSubParagraphStar}[1]{\oldsubparagraph*{#1}\mbox{}}
  \newcommand{\xxxSubParagraphNoStar}[1]{\oldsubparagraph{#1}\mbox{}}
\fi
\makeatother

\usepackage{longtable,booktabs,array}
\usepackage{calc}
\usepackage{etoolbox}
\makeatletter
\patchcmd\longtable{\par}{\if@noskipsec\mbox{}\fi\par}{}{}
\makeatother

\IfFileExists{footnotehyper.sty}{\usepackage{footnotehyper}}{\usepackage{footnote}}
\makesavenoteenv{longtable}
\usepackage{graphicx}
\makeatletter
\def\maxwidth{\ifdim\Gin@nat@width>\linewidth\linewidth\else\Gin@nat@width\fi}
\def\maxheight{\ifdim\Gin@nat@height>\textheight\textheight\else\Gin@nat@height\fi}
\makeatother

\setkeys{Gin}{width=\maxwidth,height=\maxheight,keepaspectratio}
\makeatletter
\def\fps@figure{htbp}
\makeatother

\makeatletter
\@ifpackageloaded{caption}{}{\usepackage{caption}}
\AtBeginDocument{%
\ifdefined\contentsname
  \renewcommand*\contentsname{Table of contents}
\else
  \newcommand\contentsname{Table of contents}
\fi
\ifdefined\listfigurename
  \renewcommand*\listfigurename{List of Figures}
\else
  \newcommand\listfigurename{List of Figures}
\fi
\ifdefined\listtablename
  \renewcommand*\listtablename{List of Tables}
\else
  \newcommand\listtablename{List of Tables}
\fi
\ifdefined\figurename
  \renewcommand*\figurename{Figure}
\else
  \newcommand\figurename{Figure}
\fi
\ifdefined\tablename
  \renewcommand*\tablename{Table}
\else
  \newcommand\tablename{Table}
\fi
}
\@ifpackageloaded{float}{}{\usepackage{float}}
\floatstyle{ruled}
\@ifundefined{c@chapter}{\newfloat{codelisting}{h}{lop}}{\newfloat{codelisting}{h}{lop}[chapter]}
\floatname{codelisting}{Listing}

\makeatother
\makeatletter
\@ifpackageloaded{caption}{}{\usepackage{caption}}
\@ifpackageloaded{subcaption}{}{\usepackage{subcaption}}
\makeatother

\ifLuaTeX
  \usepackage{selnolig}
\fi
\usepackage{bookmark}

\usepackage[authoryear]{natbib}  

\IfFileExists{xurl.sty}{\usepackage{xurl}}{} 
\hypersetup{
  pdftitle={Title},
  pdfauthor={Author 1; Author 2},
  pdfkeywords={3 to 6 keywords, that do not appear in the title},
  colorlinks=true,
  linkcolor={blue},
  filecolor={Maroon},
  citecolor={Blue},
  urlcolor={Blue},
  pdfcreator={LaTeX via pandoc}}

\DeclareMathOperator*{\argmin}{arg\,min}
\DeclareMathOperator*{\argmax}{arg\,max}
\DeclareMathOperator*{\argopt}{arg\,opt}

\newcommand{\anon}{1}

\begin{document}

\def\spacingset#1{\renewcommand{\baselinestretch}%
{#1}\small\normalsize} \spacingset{1}


\if1\anon
{
  \title{\textbf{SWAP Regression Methodology for Predicting Relationship with Historical Bivariate Data}}
  \author{Viral Chitlangia\thanks{Email : viralc22@iitk.ac.in}\hspace{.2cm}\\
    Department of Mathematics and Statistics, Indian Institute of
Technology Kanpur, INDIA\\
    and \\
    Mosuk Chow\thanks{Email : mxc18@psu.edu} \\
    Department of Statistics, Pennsylvania State University, State
College, Pennsylvania, USA\\
    and \\
    Sharmishtha Mitra\thanks{Email : smitra@iitk.ac.in, Corresponding author} \\
    Department of Mathematics and Statistics, Indian Institute of
Technology Kanpur, INDIA}
  \maketitle
} \fi

\if0\anon
{
  \bigskip
  \bigskip
  \bigskip
  \begin{center}
    {\LARGE\bf Title}
\end{center}
  \medskip
} \fi

\bigskip
\begin{abstract}
This study revisits regression for samples with alternating predictors (SWAP) proposed in \cite{a338225d-1414-3978-8bdb-399fb0f36185} with the purpose of finding the best fit model when the role of the response and the explanatory variables was established. In the current work, we explore the directional relationship between the two variables at a given point of time, by a novel approach which draws direct inspiration from the concept of SWAP regression. Our method, based on the Gaussian Mixture Model (GMM) and the beta distribution, while estimating the probability of a latent variable, predicts the suitable model, i.e., earmarks if a variable can take the role of an explanatory or response, at any point of time. To make this switch-over role between variables, a valid consideration, we have established the existence of a bi-directional (Granger) causality between the two variables. A detailed real data analysis of the methodology is carried out using the historical quarterly data on probably the two most intertwined macro-economic indicators explaining the health of an economy, viz., the Gross Domestic Product (GDP) and Public Debt, thereby making the application, in real data, more challenging. In particular, we use data of the US economy during the sample period 1966-2023. 
\end{abstract}

\noindent%
\vfill

\newpage
\spacingset{1.8}

\section{Introduction}
The SWAP regression method \cite{a338225d-1414-3978-8bdb-399fb0f36185} begins with a fundamental principle of bifurcation of the response-explanatory role between two variables under study. The introductory study, and so far, the only one in the literature, was motivated by a "kinesiology experiment concerning human tolerance to temperature and water vapor pressure". This was essentially an experimental setup. We further quote the authors, ".....The problem differs from traditional regression because, for one part of the data, temperature is held fixed while pressure is raised to an equilibrium point; for the other part of the data, pressure is held fixed while temperature is raised to an equilibrium point. ... Traditional regression is inadequate for modeling this type of data, because the roles of predictor and response alternate". For our work, we consider observable historical real data, thus expanding the applicability and scope of this unique regression methodology. To justify the application of SWAP we choose two highly intertwined macro-economic indicators, viz., the Gross Domestic Product (GDP) and Public Debt of an economy. We check the existence of bi-directional Granger-Causality (\cite{granger}) between the two variables. In addition, by estimating probability of a latent variable, along the lines of how Chow. \textit{et. al.} in \cite{a338225d-1414-3978-8bdb-399fb0f36185} discusses SWAP regression, we effectively settle an analogous query in our backdrop, viz., does GDP explain Public Debt, or Public Debt explain GDP; and at which time points the phenomenon of this switching over happens. A recent IMF study on public debt and real GDP (\cite{IMF}) elaborated on the role of the impact of debt on some major macro-economic indicators of the real sector (for some further reading on the topic, please refer to \cite{Domar1944}; \cite{Barro1988}; \cite{Bernheim1987}). While it is common knowledge that understanding the impact of an increasing debt on real GDP is crucial, but this assessment remains incomplete without studying the impact of GDP on debt, i.e., unless we take an objective approach to investigate the bi-directional cause and effect relationship between these two indicators. 
The SWAP regression adds a new dimension to a regression model as it can address a situation when there is no single correct answer to this query, i.e., for some parts of the sample period, when Public Debt drives GDP, the former takes the role of the predictor, and vice versa for some other parts of the sample period.

\cite{a338225d-1414-3978-8bdb-399fb0f36185} introduced a novel loss function for solving a regression model with switching roles of variables under their study, viz., pressure and temperature, say, $X$ and $Y$, with a third (binary) variable, say, $Z$, indicating which is the response, and which one the explanatory variable. 
Thus, when $Z = 0$, then $Y$ is the explanatory variable, and when $Z = 1$, $X$ is the explanatory variable. Hence, if the sample space of $(X, Y, Z)$ is the Cartesian product of $\Omega_X$ x $\Omega_Y$ x $\{$0, 1$\}$, and 
\begin{subequations}
\begin{align}
E(Y | X, Z = 1) &= g(X) \\
E(X | Y, Z = 0) &= g^{-1}(Y) \\
\mathcal{g} = \{f \in L^2(P_X) : f \text{ is a bijection}\text{ from }&\Omega_X \text{ to }\Omega_Y{, 
 }f^{-1} \in L^2(P_Y)\}, \notag
\end{align}
\end{subequations}

the final function was found by solving the following loss function on Q: $\mathcal{g} \rightarrow \mathbb{R}$:
\begin{align}
    \text{Q(}f\text{)} = E[(Y - f(X))^2 I[Z = 1]] + E[(X - f^{-1}(Y))^2 I[Z = 0]].
\end{align}
In this study we use a Gaussian Mixture Model (GMM) type representation (\cite{bishop2006pattern}) of the parametric probability model to explain the bivariate data as a weighted sum of Gaussian component densities, with unknown weight function. GMMs are widely used to model continuous variables or features in economic, financial or biometric systems. Additionally, we implement the SWAP regression assuming a beta prior for the unknown conditional probability of the indicator given the bivariate data. This is the latent variable in our case. For training the data, the Alternating-Optimization Algorithm(ALT-OPT) (\cite{inproceedings}), a technique for optimizing functions iteratively, in presence of several variables, has been effectively used here. The ALT-OPT forms the basis of the $k$-means clustering, as well as the expectation-maximization (EM) algorithm for Gaussian mixture decomposition. For optimizing a function f($x_1, x_2,..., x_n$), we use the ALT-OPT Algorithm \ref{alg:ALT-OPT} as follows:

\begin{algorithm}
    \caption{ALT-OPT Algorithm}
    \label{alg:ALT-OPT}
    \begin{algorithmic}[1]
    \item We are given n as number of observations, with ${x_1, x_2,...,x_n}$ being the n observations we are looking to optimize with respect to the function $f$.
    \item Initialize ${x_1, x_2,...,x_n}$ as $x_1^{(0)}, x_2^{(0)},..., x_n^{(0)}$
    \item For i = 1 to n
    \newline
    $x_i^{(r + 1)}$ = $\argopt_{x_i}{f(x_1^{(r + 1)},..., x_{i - 1}^{(r + 1)},x_i, x_{i + 1}^{(r)}, ..., x_n^{(r)})}$, where $\argopt$ optimizes the value of $x_i$ according to the now one variable function.
    \end{algorithmic}
\end{algorithm}

\section{SWAP Regression in a GMM backdrop}
\subsection{Setting up the Variables}
In a similar setting as \cite{a338225d-1414-3978-8bdb-399fb0f36185}, we consider bivariate data and a binary indicator variable and denote $X_0, Y_0$ as the points where Z = 0, and $X_1, Y_1$ where Z = 1. Furthermore, assume for i = $1, 2,..., n$,
\newline
if $Z_i = 1$, then $Y_{1i} = g(X_{1i}) + \epsilon_{1i}$, where $\epsilon_{1i} \sim N(0, \sigma_{1}^2)$, and 
\newline
if $Z_i = 0$, then $X_{0i} = g^{-1}(Y_{0i}) + \epsilon_{0i}$, where $\epsilon_{0i} \sim N(0, \sigma_{0}^2)$.
\newline
Note that here we have the presence of a latent variable Z. If we had known Z apriori, we could easily obtain the likelihood function using
\begin{subequations}
\begin{align}
f(x, y | z = 0) &= \frac{1}{\sqrt{2\pi \sigma_{0}^2}} e^{-\frac{(x - g^{-1}(y))^2}{2 \sigma_0^2}} f(y | z = 0) \\
f(x, y | z = 1) &= \frac{1}{\sqrt{2\pi \sigma_{1}^2}} e^{-\frac{(y - g(x))^2}{2 \sigma_1^2}} f(x | z = 1)
\end{align}
\end{subequations}

\subsection{Relevance to the GMM }
Although the problem statement looks like a GMM representation, this is not exactly in the form of the GMMs, where we have constant means for all the Gaussian distributions (\cite{bishop2006pattern}).
We need to keep in mind that the "mean" of the first distribution is not independent of the "mean" of the second distribution.
Our idea to solve this problem is by implementing the ALT-OPT algorithm (\cite{inproceedings}). We have a total of six parameters to predict - ($\pi_0, g, Z, \sigma_0^2, \sigma_1^2$, $n_0$). We alternatively optimize for $\{\pi_0, \pi_1, Z, n_0, n_1\}$ and $\{g, \sigma_0^2, \sigma_1^2\}$.
Here, $Z_i \sim Bernoulli(\pi_1)$ and $\pi_1 = 1 - \pi_0$. In addition, $n_1 = n - n_0$, where $n_0$ is the number of observations with $Y$ as the explanatory variable, and $n_1$ is the number of observations with $X$ as the explanatory variable. $\pi_0$ is the prior probability of a random observation having $Y$ as the explanatory variable, and $\pi_1$ is the prior probability of a random observation having $X$ as the explanatory variable.

\subsubsection{Proposed Algorithm}
Let us initialize $Z^{(0)}$ randomly. We can also initialize $g^{(0)}$ as any suitable model for $X$ on $Y$. Using that, we can get the vectors $\epsilon_0^{(0)} = X_0^{(0)} - {g^{(0)}}^{-1}(Y_0^{(0)})$ and $\epsilon_1^{(0)} = Y_1^{(0)} - g^{(0)}(X_1^{(0)})$. Obtaining the maximum likelihood estimate (MLE) of the variance of errors $\epsilon_0^{(0)}$ and $\epsilon_1^{(0)}$, we can initialize $\sigma_0^{2(0)}$ and $\sigma_1^{2(0)}$. $n_0^{(0)} = \sum_{i=1}^n I[Z_i^{(0)} = 0]$ and $n_1^{(0)} = n - n_0^{(0)}$. We get $\pi_0^{(0)} = \frac{n_0^{(0)}}{n}$ and $\pi_1^{(0)} = 1 - \pi_0^{(0)}$.

\begin{enumerate}
\item \textbf{Iterations}
\newline
Here, we assume that the prior distributions of $X$ and $Y$ are independent of $Z$. Let us denote the density functions by $h_X$ and $h_Y$ for $X$ and $Y$, respectively. Thus,

$P(Z = 1 | X, Y) \propto f(X, Y | Z = 1) P(Z = 1) \propto f(Y | X, Z = 1) P(Z = 1) P(X) \propto \pi_1 \frac{1}{\sqrt{2\pi \sigma_{1}^2}} e^{-\frac{(Y - g(X))^2}{2 \sigma_1^{2}}} h_X(X)$, and

$P(Z = 0 | X, Y) \propto f(X, Y | Z = 0) P(Z = 0) \propto f(X | Y, Z = 0) P(Z = 0) P(Y) \propto \pi_0 \frac{1}{\sqrt{2\pi \sigma_0^2}} e^{-\frac{(X - g^{-1}(Y))^2}{2 \sigma_0^{2}}} h_Y(Y)$.
\newline
\newline
In exact form,
\newline
\begin{subequations}
\begin{align}
\label{eq:1}
P(z^{(t + 1)} = 1 | x, y) &= \frac{\pi_1^{(t)} \frac{1}{\sqrt{2\pi \sigma_{1}^{2(t)}}} e^{-\frac{(y - g^{(t)}(x))^2}{2 \sigma_1^{2(t)}}} h_X(x)}{\pi_1^{(t)} \frac{1}{\sqrt{2\pi \sigma_{1}^{2(t)}}} e^{-\frac{(y - g^{(t)}(x))^2}{2 \sigma_1^{2(t)}}} h_X(x) + \pi_0^{(t)} \frac{1}{\sqrt{2\pi \sigma_{0}^{2(t)}}} e^{-\frac{(x - {g^{(t)}}^{-1}(y))^2}{2 \sigma_0^{2(t)}}} h_Y(y)} \\
\label{eq:2}
P(z^{(t + 1)} = 0 | x, y) &= \frac{\pi_0^{(t)} \frac{1}{\sqrt{2\pi \sigma_{0}^{2(t)}}} e^{-\frac{(x - {g^{(t)}}^{-1}(y))^2}{2 \sigma_0^{2(t)}}} h_Y(y)}{\pi_1^{(t)} \frac{1}{\sqrt{2\pi \sigma_{1}^{2(t)}}} e^{-\frac{(y - g^{(t)}(x))^2}{2 \sigma_1^{2(t)}}} h_X(x) + \pi_0^{(t)} \frac{1}{\sqrt{2\pi \sigma_{0}^{2(t)}}} e^{-\frac{(x - {g^{(t)}}^{-1}(y))^2}{2 \sigma_0^{2(t)}}} h_Y(y)}
\end{align}
\end{subequations}
\newline
\newline
We have the following conditional likelihood function 
\newline
\begin{equation}
\label{eq:3}
L(\theta | X, Y, Z^{(t + 1)}, \theta^{(t)}) = \prod_{i = 1}^{n_0^{(t)}} \frac{1}{\sqrt{2\pi \sigma_{0}^{2(t)}}} e^{-\frac{(X_{0i} - g^{-1}(Y_{0i}))^2}{2 \sigma_0^{2(t)}}} h_Y(Y_{0i}) \prod_{i = 1}^{n_1^{(t)}} \frac{1}{\sqrt{2\pi \sigma_{1}^{2(t)}}} e^{-\frac{(Y_{1i} - g(X_{1i}))^2}{2 \sigma_1^{2(t)}}} h_X(X_{1i})
\end{equation}
\newline
We then get $g^{(t + 1)}$ = $\argmax_g$ L($\theta$ $|$ X, Y, $Z^{(t + 1)}$, $\theta^{(t)}$).
\newline
\begin{algorithm}
    \caption{New GMM Algorithm}
    \label{alg:GMM}
    \begin{algorithmic}[1]
    \item For all $Z_i$, we can predict \newline $Z_i^{(t + 1)} = \argmax({P(Z_i^{(t + 1)} = 0 | X_i, Y_i), P(Z_i^{(t + 1)} = 1 | X_i, Y_i)})$, using equation \ref{eq:1} and equation \ref{eq:2}.
    \item $g^{(t + 1)}$ = $\argmin_g$ $\sum_{i = 1}^{n} (\frac{(Y_i - g(X_i))^2}{\sigma_1^{2(t)}} I[Z_i^{(t + 1)} = 1] + \frac{(X_i - g^{-1}(Y_i))^2}{\sigma_0^{2(t)}} I[Z_i^{(t + 1)} = 0])$, using equation \ref{eq:3}.
    \item $\epsilon_0^{(t + 1)} = X_0^{(t + 1)} - {g^{(t + 1)}}^{-1}(Y_0^{(t + 1)})$.
    \item $\epsilon_1^{(t + 1)} = Y_1^{(t + 1)} - g^{(t + 1)}(X_1^{(t + 1)})$. 
    \item $\sigma_0^{2(t + 1)}$ = Var($\epsilon_0^{(t + 1)}$).
    \item $\sigma_1^{2(t + 1)}$ = Var($\epsilon_1^{(t + 1)}$).
    \item $n_0^{(t + 1)} = \sum_i^n I[Z_i^{(t + 1)} = 0].$
    \item $n_1^{(t + 1)} = \sum_i^n I[Z_i^{(t + 1)} = 1].$
    \end{algorithmic}
\end{algorithm}

\item \textbf{Stopping Criterion}
\newline
Stop when
\begin{itemize}
  \item $Z^{(t)} = Z^{(t + 1)}$.
  \item $||g^{(t + 1)} - g^{(t)}|| < \epsilon$
  \item $\text{iterations} > \text{max}\_\text{iters}$
\end{itemize}
\end{enumerate}

\subsection{Relevance to the Beta Distribution}
We assume the distribution of $\pi_i = P(Z_i^{(t + 1)} = 0 | X_i, Y_i)$ as a random draw from a Beta distribution, say, $Beta(\alpha, \beta)$, where $0<\alpha, \beta <1$ are unknown parameters of the distribution. Now, since we want the values of $\pi_i$ to be far from 0.5, or close to 0 or 1, ideally we would like the parameters of the Beta distribution as low as possible. To estimate the parameters, we use the method of maximum likelihood estimation.
\subsubsection{Transformation of the Beta Function}
To minimize $\alpha$ and $\beta$, we would like to have a closed form expression of $\hat{\alpha}_{MLE}$ and $\hat{\beta}_{MLE}$. For a beta distribution, this is not possible. \cite{volodin1970beta} showed that the cumulative distribution function (cdf) of a beta distribution can be approximated as $J_x(\alpha, \beta)$, where $\alpha$ and $\beta$ are small. Since we want $\alpha$ and $\beta$ to be small, we use $J_x$ to approximate the beta cdf as follows:

\begin{subequations}
\begin{align}
J_x(\alpha, \beta) &=
    \begin{cases}
        (1 - \gamma)(\frac{x}{1 - x})^\alpha & \text{if } 0 \leq x \leq 0.5\\
        1 - \gamma(\frac{1 - x}{x})^\beta & \text{if } 0.5 \leq x \leq 1
    \end{cases} \\
\epsilon &= \alpha + \beta \\
\gamma &= \frac{\alpha}{\epsilon}
\end{align}
\end{subequations}

Using this, we can get $j(x;\alpha, \beta) = \frac{d}{dx} J_x(\alpha, \beta)$.
\begin{equation}
j(x ; \alpha, \beta) =
    \begin{cases}
        (1 - \gamma) \alpha (\frac{x}{1 - x})^{\alpha - 1} \frac{1}{(1 - x)^2} & \text{if } 0 \leq x \leq 0.5\\
        \gamma \beta (\frac{1 - x}{x})^{\beta - 1} \frac{1}{x^2} & \text{if } 0.5 \leq x \leq 1
    \end{cases}
\end{equation}

\subsubsection{Maximum Likelihood Estimates for $\alpha$ and $\beta$}
Suppose that we have $X_i \sim J(\alpha, \beta)$, for i = 1 to n. We have $X_0$ = X, where X $\leq$ 0.5, and $X_1$ where X $\geq$ 0.5. Define $\pi_1 = \frac{n_1}{n_0 + n_1} = \frac{1}{n}\sum_{i=1}^n I(X_i > 0.5)$.
\begin{equation}
    E[\pi_1] = \frac{1}{n}\sum_{i=1}^n P(X_i > 0.5)
\end{equation}
Since $X_i$ are identically distributed, we can write as -
\begin{equation}
    E[\pi_1] = P(X_1 > 0.5)
\end{equation}
Since, $X_i ~ \sim Beta(\alpha, \beta)$, we can approximate it as $J_x(\alpha, \beta)$.
\begin{subequations}
\begin{align}
    E[\pi_1] &= \int_{0.5}^{1} j(x;\alpha, \beta) dx \\
    E[\pi_1] &= J_{1}(\alpha, \beta) - J_{0.5}(\alpha, \beta) \\
    E[\pi_1] &= \gamma
\end{align}
\end{subequations}
Since, $\pi_1$ is an unbiased estimator for $\gamma$, we can replace $\gamma$ with $\pi_1$. As we will see later on, this makes computations a lot easier.
\begin{subequations}
\begin{align}
    log \ L(\alpha, \beta \ | \ X) &= \sum_{i = 1}^{n_0} [log(\frac{1 - \gamma}{(1 - X_{0i})^2}) + log(\alpha) + (\alpha - 1)log(\frac{X_{0i}}{1 - X_{0i}})] \notag\\
    &+ \sum_{i = 1}^{n_1} [log(\frac{\gamma}{X_{1i}^2}) + log(\beta) + (\beta - 1)log(\frac{1 - X_{1i}}{X_{1i}})] \notag
\end{align}
\end{subequations}
Maximizing it wrt $\alpha$ and $\beta$, respectively, yield 
\begin{equation} 
\label{eq:4}
    \hat{\alpha} = \frac{n_0}{\sum_{i = 1}^{n_0} log(\frac{1 - X_{0i}}{X_{0i}})}
\end{equation}
and
\begin{equation}
\label{eq:5}
    \hat{\beta} = \frac{n_1}{\sum_{i = 1}^{n_1} log(\frac{X_{1i}}{1 - X_{1i}})}
\end{equation}

\subsubsection{Loss Function}
Since we want to have the minimum values of $\hat{\alpha}$ and $\hat{\beta}$ obtained in equation \ref{eq:4} and equation \ref{eq:5}, we can define the new loss function as -
\begin{subequations}
\begin{align}
\label{eq:6}
    L(\theta | X) &= -\frac{n_0}{\hat{\alpha}} - \frac{n_1}{\hat{\beta}} \\
    L(\theta|X) &= -\sum_{i = 1}^{n_0} log(\frac{1 - X_{0i}}{X_{0i}}) - \sum_{i = 1}^{n_1} log(\frac{X_{1i}}{1 - X_{1i}})
\end{align}
\end{subequations}

We can make sense of equation \ref{eq:6} by the fact that minimizing it will minimize the parameters of the beta distribution, and we are regularizing the values of $\alpha$ and $\beta$ using the number of values of observation to determine them. Any loss function could have been chosen, which minimizes the parameters. The main reason we choose this particular loss function is because of its similarities to the one we got using GMM. Ignoring the constants, we get - 
\begin{equation}
\label{eq:7}
    L(\theta | X, Y) = \sum_{i = 1}^{n_0} [\frac{(X_{0i} - g^{-1}(Y_{0i}))^2}{\sigma_0^2} - \frac{(Y_{0i} - g(X_{0i}))^2}{\sigma_1^2}] + \sum_{i = 1}^{n_1} [\frac{(Y_{1i} - g(X_{1i}))^2}{\sigma_1^2} - \frac{(X_{1i} - g^{-1}(Y_{1i}))^2}{\sigma_0^2}]
\end{equation}
We get the final loss function as equation \ref{eq:7}. We can make sense of it as the logarithm of equation \ref{eq:3}, but with a regularization term, to try to separate Z = 0 from Z = 1, and Z = 1 from Z = 0. 

\subsubsection{Proposed Algorithm}
We can use Algorithm \ref{alg:GMM}, to optimize using the new loss function \ref{eq:7}, as shown in Algorithm \ref{alg:Beta}.

\begin{algorithm}
    \caption{Beta SWAP Algorithm}
    \label{alg:Beta}
    \begin{algorithmic}[1]
    \item For all $Z_i$, we can predict \newline $Z_i^{(t + 1)} = \argmax({P(Z_i^{(t + 1)} = 0 | X_i, Y_i), P(Z_i^{(t + 1)} = 1 | X_i, Y_i)})$, using equation \ref{eq:1} and equation \ref{eq:2}.
    \item $g^{(t + 1)}$ = $\argmin_g$ $\sum_{i = 1}^{n} [\frac{(Y_{i} - g(X_{i}))^2}{\sigma_1^{2(t)}} - \frac{(X_{i} - g^{-1}(Y_{i}))^2}{\sigma_0^{2(t)}}]I[Z_i^{(t + 1)} = 1] + [\frac{(X_{i} - g^{-1}(Y_{i}))^2}{\sigma_0^{2(t)}} - \frac{(Y_{i} - g(X_{i}))^2}{\sigma_1^{2(t)}}] I[Z_i^{(t + 1)} = 0])$, using equation \ref{eq:7}.
    \item $\epsilon_0^{(t + 1)} = X_0^{(t + 1)} - {g^{(t + 1)}}^{-1}(Y_0^{(t + 1)})$.
    \item $\epsilon_1^{(t + 1)} = Y_1^{(t + 1)} - g^{(t + 1)}(X_1^{(t + 1)})$. 
    \item $\sigma_0^{2(t + 1)}$ = Var($\epsilon_0^{(t + 1)}$).
    \item $\sigma_1^{2(t + 1)}$ = Var($\epsilon_1^{(t + 1)}$).
    \item $n_0^{(t + 1)} = \sum_i^n I[Z_i^{(t + 1)} = 0].$
    \item $n_1^{(t + 1)} = \sum_i^n I[Z_i^{(t + 1)} = 1].$
    \end{algorithmic}
\end{algorithm}

\section{Data Analysis}
\subsection{Data Source and Description}
Our idea is to implement SWAP Regression for identifying any bi-directional causal data points arising in the wide plethora of macro and micro economic, financial, social, bio-medical and health indicators. In this discussion, we limit ourselves to macro-economic indicators. \cite{DEVITA201855} concluded that Public Debt and GDP growth are bi-directional causal for the US market. In our research, we consider US Public Debt and Nominal GDP, both in the form of stock variables, and not in growth, ratio, or rate form. The data source is Kaggle, \href{https://www.kaggle.com/datasets/pavankrishnanarne/us-public-debt-quarterly-data-1996-present}{US Public Debt Quarterly Data(in Billion \$)} and \href{https://fred.stlouisfed.org/series/NGDPSAXDCUSQ}{US Nominal GDP Quarterly Data(in Million \$)}, in quarters. Public Debt data was available from 1966-01-01 to 2023-01-01, and Nominal GDP data from 1966-01-01 to 2023-01-01. We have scaled the gdp and public debt data by a factor of $10^{-6}$.

\subsection{Testing Bi-directional Granger Causality}
To verify our claim that the two indicators are, in fact, a good choice of applicability in SWAP regression set-up, we test bi-directional causality, by the traditional econometric Granger Causality test (\cite{702ab909-8cb1-3c30-a5f1-ab4517d6cf1c}) using the \textbf{grangertest} function in vector autoregressive (VAR) set-up from the \textbf{vars} package (\cite{VAR1} \& \cite{VAR2}) in R.
\newline
Here, $X$ represents Nominal GDP, and $Y$ represents Public Debt.
\begin{figure}[htp]
\centering
\includegraphics[width=9cm]{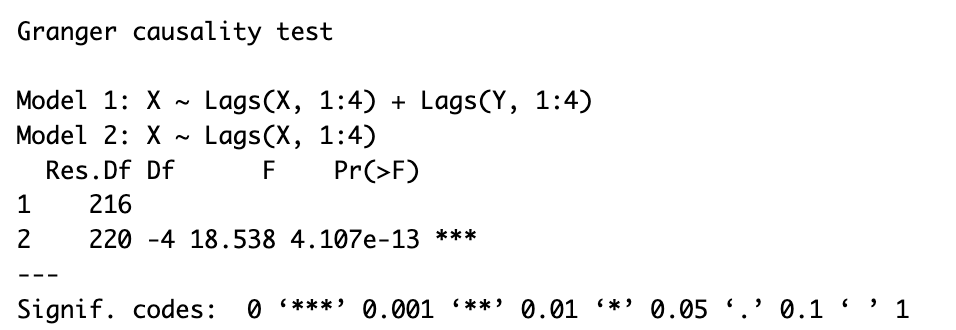}
\centering
\caption{Granger Test for Y causing X}
\end{figure}
\begin{figure}[htp]
\centering
\includegraphics[width=9cm]{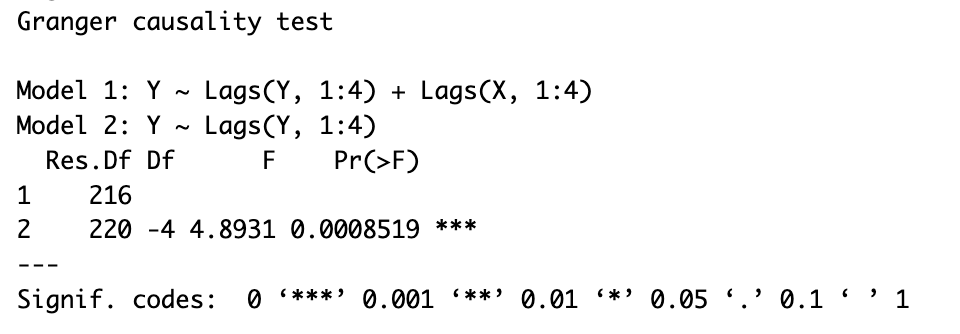}
\centering
\caption{Granger Test for X causing Y}
\end{figure}

Fitting the data showed the significance of each indicator in the respective model.
\begin{figure}[htp]
\centering
\includegraphics[width=9cm]{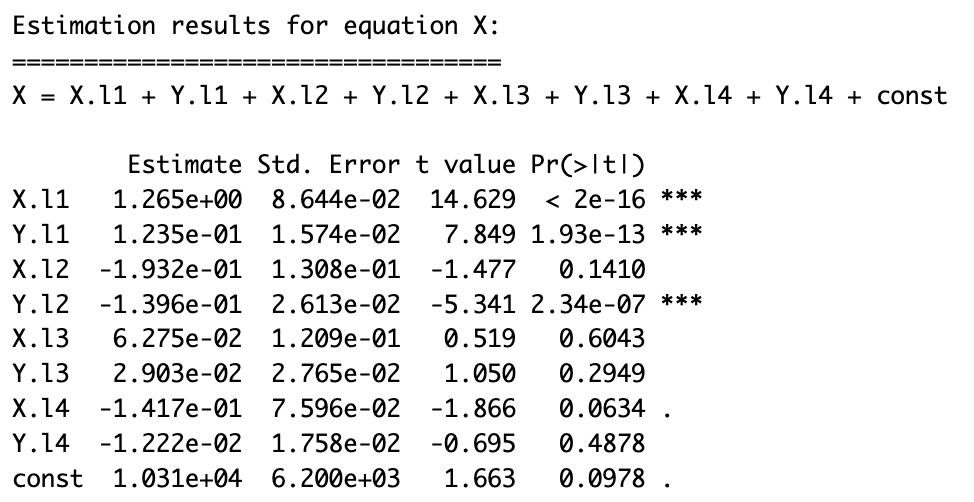}
\centering
\caption{Significance of Y in predicting X}
\end{figure}
\begin{figure}[htp]
\centering
\includegraphics[width=9cm]{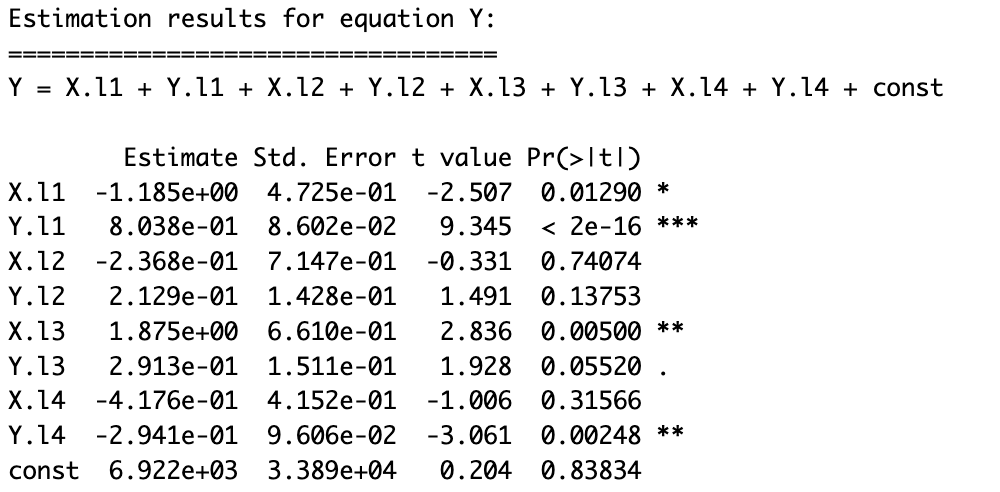}
\centering
\caption{Significance of X in predicting Y}
\end{figure}

Since for the models trained on GDP, we can find a lag of Public Debt, which is significant, and vice versa, we can conclude bi-directional causality between them. We obtained similar results of existence of bi-directional causality between the two variables by the \textbf{sem} function from the \textbf{lavaan} package (\cite{lavaan}) in R to test bi directional causality using structural equation models (SEM). We also got similar results using the \textbf{systemfit} function from the \textbf{systemfit} package (\cite{systemfit}) in R to test bi directional causality using simultaneous equation model (SIM).

\subsection{Testing Stationarity in Residuals}
Since we are essentially in the regression model set-up, care should be taken to check we are not running into spurious regression in our attempt to verify the bi-directional causality. For this purpose, we check whether residuals from the fits are stationary or not. We got the residuals after fitting the models for the Granger Causality Test and hence apply the Augmented Dickey Fuller Test (\cite{HARRIS1992381}) to test stationarity of the residuals.
The residuals of both the models gave a p-value of less than 0.01 on Augmented Dickey Fuller test, which we applied using the \textbf{adf.test} function from the \textbf{aTSA} package (\cite{aTSA}) in R.

\subsection{Theoretical distributions of the economic variables under study }
We use the SWAP regression model with the US GDP and Public Debt data.
To implement the model, we assume theoretical distributions of $X$ and $Y$.

\begin{figure}
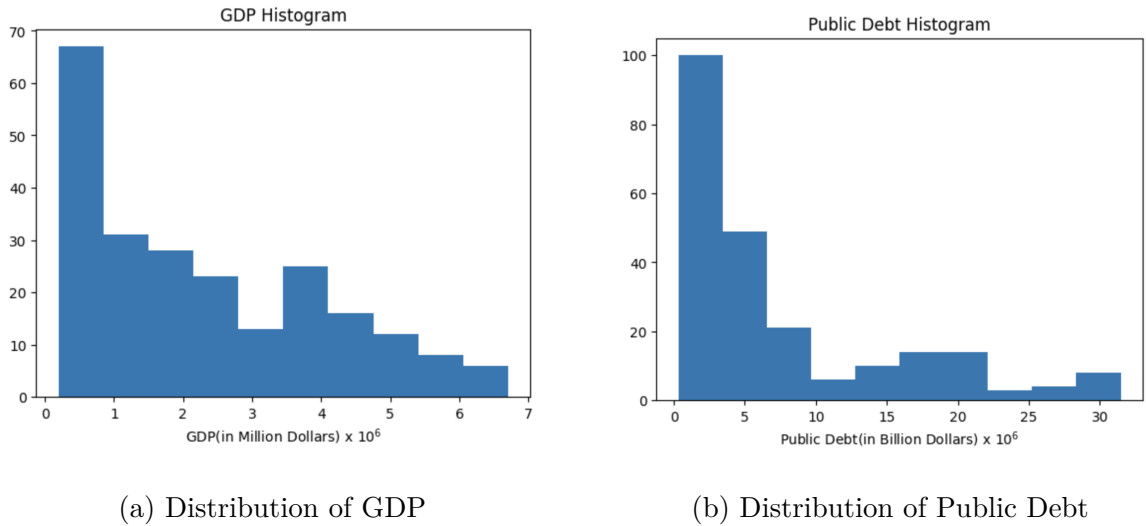

\begin{subfigure}{.5\textwidth}
\centering
\includegraphics[width=7.5cm]{XHist.pdf}
\caption{Distribution of GDP}
\label{fig:XHist}
\centering
\end{subfigure}
\hfill
\begin{subfigure}{.5\textwidth}
\centering
\includegraphics[width=7.5cm]{YHist.pdf}
\caption{Distribution of Public Debt}
\label{fig:YHist}
\centering
\end{subfigure}
\caption{Distribution of Data}
\end{figure}

To assume a distribution on US GDP and Public Debt, we look at the histogram of US GDP, as shown in Figure \ref{fig:XHist} and that of Public Debt, as shown in Figure \ref{fig:YHist}. We can see that both resemble exponential distribution with rate parameter, say, $\lambda$. We can confirm the choice of the distribution by the univariate Kolmogorov-Smirnov test (\cite{Massey01031951}), and use the kstest function in the scipy.stats library (\cite{2020SciPy-NMeth}) for this purpose. For GDP, it gives a p-value of $9.754 \times 10^{-26}$, and for Public Debt, it gives a p-value of $1.278 \times 10^{-61}$ We obtain maximum likelihood estimates of the rate parameter $\hat{\lambda}_{GDP} = \frac{1}{\bar{X}} = 0.4387$ and $\hat{\lambda}_{Debt} = \frac{1}{\bar{Y}} = 0.1349$ for GDP and Debt, respectively.

\section{Fitting the Model}
We fit a total of six models with $X$ as GDP(in Million Dollars) $\times$ $10^6$ and $Y$ as Public Debt(in Billion Dollars) $\times$ $10^6$ of US. These are the simple linear regression (SLR), quadratic regression (QR), the SWAP GMM linear and quadratic regression, and the SWAP beta linear and quadratic regression models. 

\begin{enumerate}
\item \textbf{Simple Linear Regression}
\newline
y = 4.4972 $x$ - 2.8404

\item \textbf{Quadratic Regression}
\newline
y = 0.6460 $x^2$ + 0.8578 $x$ + 0.1158

\item \textbf{SWAP GMM Linear Regression}
\newline
y = 4.4877 $x$ - 2.8283

\item \textbf{SWAP GMM Quadratic Regression}
\newline
y = 0.6221 $x^2$ + 0.9642 $x$ + 0.0488

\item \textbf{SWAP Beta Linear Regression}
\newline
y = 4.4877 $x$ - 2.8283

\item \textbf{SWAP Beta Quadratic Regression}
\newline
y = 0.6226 $x^2$ +  0.9641 $x$ + 0.0486
\end{enumerate}

\begin{figure}[h]
\begin{subfigure}{.5\textwidth}
\centering
\includegraphics[width=7.5cm]{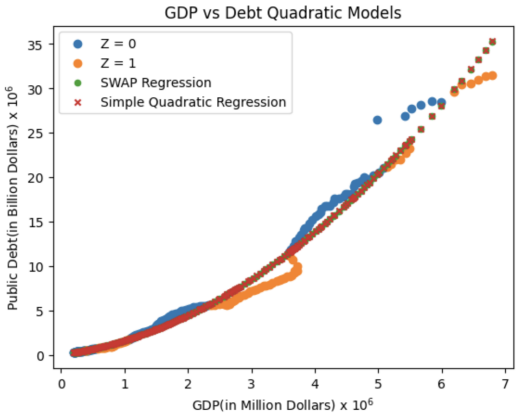}
\caption{}
\label{QM}
\centering
\end{subfigure}
\hfill
\begin{subfigure}{.5\textwidth}
\centering
\includegraphics[width=7.5cm]{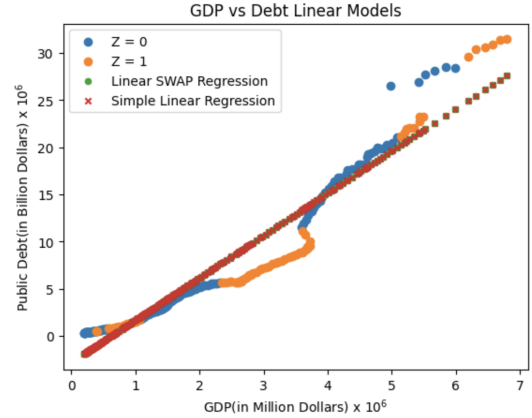}
\caption{}
\label{LM}
\centering
\end{subfigure}
\caption{Model Fitting for GMM Method}
\label{GMMGraph}
\end{figure}

\begin{figure}[h]
\begin{subfigure}{.5\textwidth}
\centering
\includegraphics[width=7.5cm]{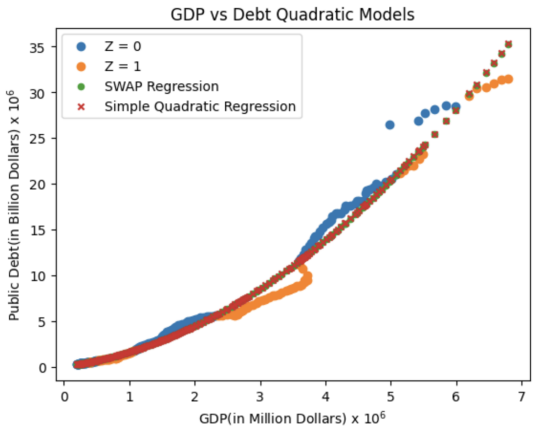}
\caption{}
\label{QBM}
\centering
\end{subfigure}
\hfill
\begin{subfigure}{.5\textwidth}
\centering
\includegraphics[width=7.5cm]{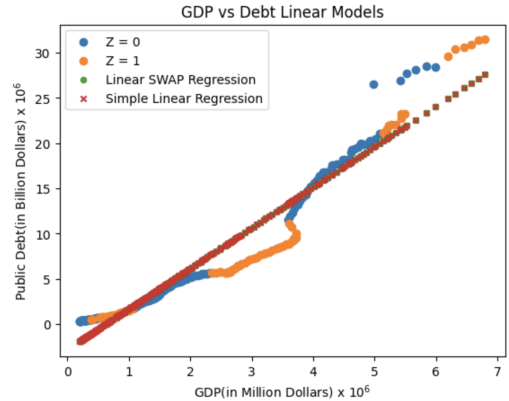}
\caption{}
\label{LBM}
\centering
\end{subfigure}
\caption{Model Fitting for Beta Method}
\label{BetaGraph}
\end{figure}
We can see the model fits in plots \ref{QM}, \ref{LM}, \ref{QBM} and \ref{LBM}. We can see the clusters of the explanatory variable forming in the plots, color coded with blue when Public Debt is the explanatory variable, and orange when GDP is the explanatory variable. Simple regression and SWAP regression for all 4 have symbols of "x" and "o" respectively to more clearly distinguish them both.

\section{Testing Goodness of Fit of SWAP}
The test of goodness of the fit can be determined by how well our method can cluster the data into Z = 0 and Z = 1. We can define a good fit as one with more "surety" of Z to be 0 or 1. For the model used, we can look at the probabilities of Z = 1, as done in equation \ref{eq:1}. We can get a histogram of the probabilities to visualize it. However, since $P_{Z|X, Y}$ is a probability, it is between 0 and 1, so we can assume it to follow a $Beta$ distribution. Since, we want that value to be close to 0 or 1, for more "surety" of whether Z is 0 or 1, we would want $\alpha$ and $\beta$ to be as less than 1 as possible.

\begin{figure}
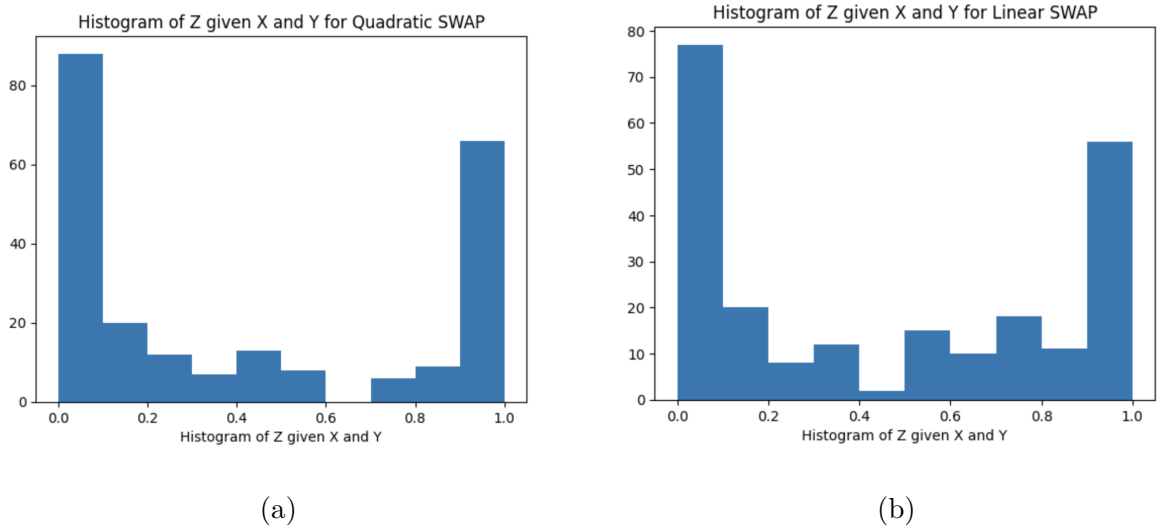

\begin{subfigure}{.5\textwidth}
\centering
\includegraphics[width=7.5cm]{HistZQuad.pdf}
\caption{}
\label{fig:ZHistQ}
\centering
\end{subfigure}
\hfill
\begin{subfigure}{.5\textwidth}
\centering
\includegraphics[width=7.5cm]{HistZLin.pdf}
\caption{}
\label{fig:ZHistL}
\centering
\end{subfigure}
\caption{Conditional Distribution of P(Z = 1) for GMM Model}
\end{figure}

\begin{figure}
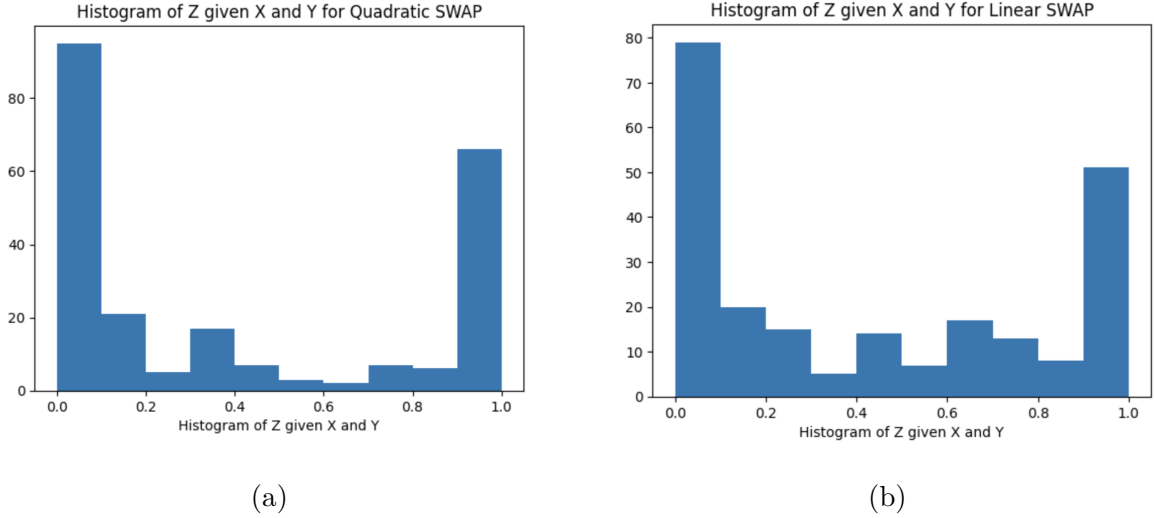

\begin{subfigure}{.5\textwidth}
\centering
\includegraphics[width=7.5cm]{HistZQuad-Beta.pdf}
\caption{}
\label{fig:ZHistQ-Beta}
\centering
\end{subfigure}
\hfill
\begin{subfigure}{.5\textwidth}
\centering
\includegraphics[width=7.5cm]{HistZLin-Beta.pdf}
\caption{}
\label{fig:ZHistL-Beta}
\centering
\end{subfigure}
\caption{Conditional Distribution of P(Z = 1) for Beta Model}
\end{figure}

The plots \ref{fig:ZHistQ}, \ref{fig:ZHistL} for GMM Model and plots \ref{fig:ZHistQ-Beta} and \ref{fig:ZHistL-Beta} show that the values of Z we got are generally closer to 0 and 1 than to 0.5. To test it mathematically, we can obtain the Maximum Likelihood Estimates for $\alpha$ and $\beta$ of both distributions.
\newline
\newline
For the Quadratic regression with GMM Modeling, we get $\hat{\alpha}_{MLE} = 0.0861$, $\hat{\beta}_{MLE} = 0.1117$.
\newline
For the Linear regression with GMM Modeling, we get $\hat{\alpha}_{MLE} = 0.2759$, $\hat{\beta}_{MLE} = 0.2848$.
\newline
\newline
For the Quadratic regression with Beta Modeling, we get $\hat{\alpha}_{MLE} = 0.0856$, $\hat{\beta}_{MLE} = 0.1128$.
\newline
For the Linear regression with Beta Modeling, we get $\hat{\alpha}_{MLE} = 0.2861$, $\hat{\beta}_{MLE} = 0.3201$.
\newline
\newline
Here, we can see that the estimates for the Quadratic SWAP are smaller than the estimates for the Linear SWAP Model, and are less than 1, in both cases, GMM and Beta, so we can say that the Quadratic SWAP Model is a good fit and is better than the Linear SWAP Model.

\section{Using SWAP to predict Causal Variable}
The final SWAP model can be used to determine when $X$ causes $Y$ and when $Y$ causes $X$. This can be determined by the distribution of $Z$.
Using both models, we can store the final values of $P_{Z^{(t)}|X, Y}$, and $Z^{(t)}$. This can be modeled with the time component to find a relation between $X$ and $Y$, and according to $Z = 0$ or $Z = 1$, we can predict the explanatory and response variable. We can explore causality at those regions.
\begin{enumerate}
    \item \textbf{GMM Model} 
\newline
\begin{figure}
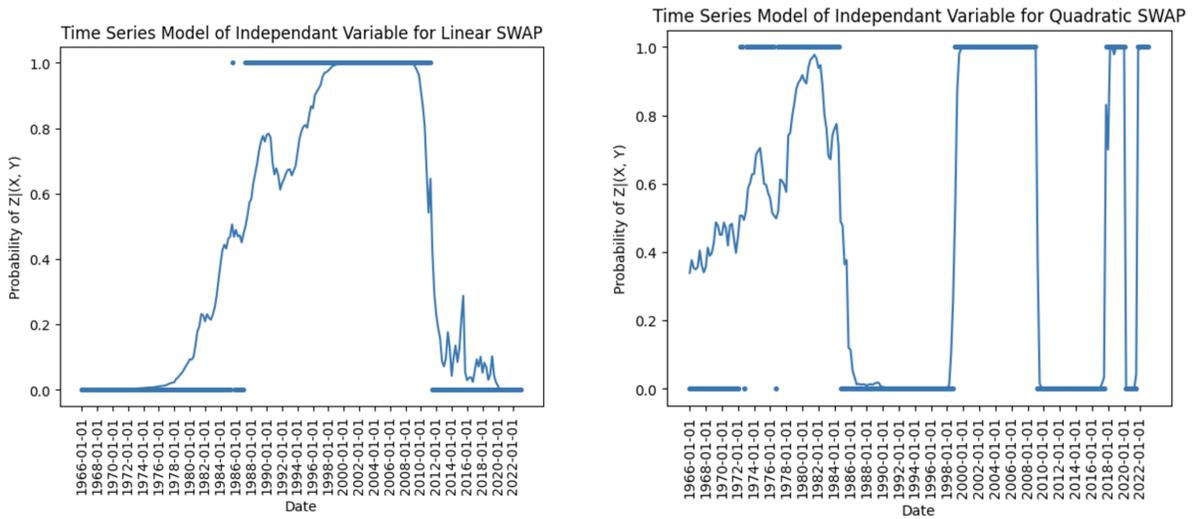

\begin{subfigure}{.5\textwidth}
\centering
\includegraphics[width=8cm]{GMMLinearPredZ.pdf}
\caption{}
\label{fig:GMMLinZ}
\centering
\end{subfigure}
\hfill
\begin{subfigure}{.5\textwidth}
\centering
\includegraphics[width=8cm]{GMMQuadraticPredZ.pdf}
\caption{}
\label{fig:GMMQuadZ}
\centering
\end{subfigure}
\caption{Swapping Variables wrt Time for GMM Model}
\label{fig:GMMZ}
\end{figure}
For the implementation of SWAP GMM to predict the causal variable, we refer to Figure \ref{fig:GMMZ}. We see that GDP drives Public Debt in the years 1988-2012 for the Linear Case, and 1972-1984, 2000-2009, 2018-2020, 2022-present. Public Debt drives the GDP in the other years. 

\item \textbf{Beta Model}
\newline
\begin{figure}
\begin{subfigure}{.5\textwidth}
\centering
\includegraphics[width=7.5cm]{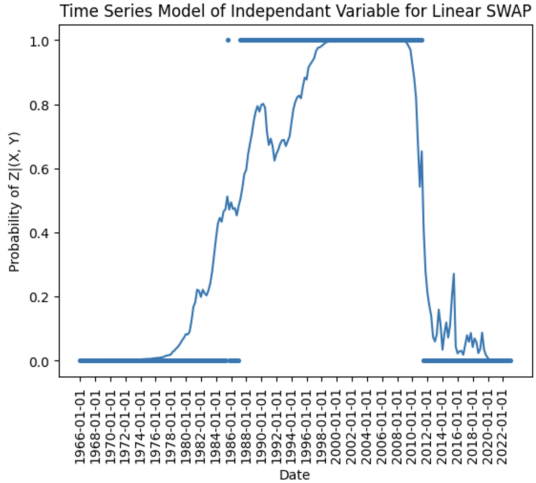}
\caption{Beta Linear}
\label{fig:BetaLinZ}
\centering
\end{subfigure}
\hfill
\begin{subfigure}{.5\textwidth}
\centering
\includegraphics[width=8cm]{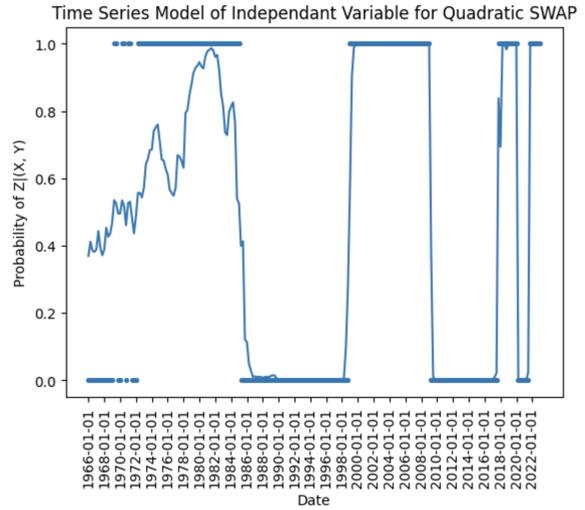}
\caption{Beta Quadratic}
\label{fig:BetaQuadZ}
\centering
\end{subfigure}
\caption{Swapping Variables wrt Time for Beta Model}
\label{fig:BetaZ}
\end{figure}
For the implementation of SWAP GMM to predict the causal variable, we refer to Figure \ref{fig:BetaZ}. We see that GDP drives Public Debt in the years 1988-2012 for the Linear Case, and 1972-1984, 2000-2009, 2018-2020, 2022-present. Public Debt drives the GDP in the other years. 
\end{enumerate}

Since both models give same years for change between the two variables, we can directly analyze them together. For the quadratic relationship case, we see that 1966-1972, 1984-2000, 2009-2018 and 2020-2022 are the times when Public Debt drives the GDP. These can be analyzed and explained in the backdrop of various historical factors having far reaching economic repercussions encompassing the real, fiscal, monetary and financial sectors among others. While fiscal and taxation policies lowered debt, renewed monetary policies increased inflation and gave a boost to GDP (\cite{Bordo2018TheIO}). Subsequent recession periods in the eighties explain the shift in the explanatory variable from GDP to Public Debt in 1984. We can also see the impact of the financial crisis of 2008  (Great Recession), and the Covid crisis of 2020-22 caused the Public Debt to drive the GDP. Post 2022, when the GDP stabilized it became the driving force. On the other hand, for the linear relationship case, we see that turning points are fewer - 1988 (Public Debt drives GDP), 2010 (roles of Public Debt and GDP switch) and, again in, 2010  with debt driving GDP. The graph does not reflect any change after that, but as already claimed by us, the quadratic relationship is much more accurate than the linear one.

\section{Conclusion}
In this paper, we use the idea of SWAP regression to choose the explanatory variable and response variable in different epochs of a given sample period. Having established a bi-directional causality between the two variables, a methodology is developed to explore potential explanatory relationship during those periods and to identify the predictor variable at each data point. A detailed real data analysis of the methodology is carried out using the historical quarterly data on probably the two most intertwined macro-economic indicators explaining the health of an economy, viz., the Gross Domestic Product (GDP) and Public Debt, thereby making the application in real data, more challenging. We implemented the methodology on the US GDP vs Public Debt relationship during the period 1966 to 2023. A finer analysis of some important historical events of economic significance, can match the changes of explanatory variables to such events. This idea could potentially be used to predict future recessions, or major upheavals in the economic conditions of the economy. We studied the phenomena through two models, viz., the GMM and the beta models. The $beta(\alpha, \beta)$ model is subjective to the choice of the loss function to minimize $\hat{\alpha}$ and $\hat{\beta}$. With our loss function, we assigned weight to each side, with respect to its data points. Regarding the latent variable driving the response-explanatory role of the two variables, when there are more instances of  Z = 1 than Z = 0, we would assign more weight to the Z = 1 cases than to the Z = 0 cases, and vice versa, according to how many such cases there are. In the quadratic relationship, we see that the beta model, which was created to reduce the proposed loss function, performed better than the GMM model. However, in the linear relationship, the GMM performs slightly better. Overall, the quadratic relationship is significantly better than the linear one. 

\section{Disclosure Statement}
None of the authors have any conflict of interests.

\section{Data Availability Statement}
In our research, we use US Public Debt and Nominal GDP, both in the form of stock variables, and not in growth, ratio, or rate form. The data source is Kaggle, US Public Debt Quarterly Data(in Billion \$) (\href{https://www.kaggle.com/datasets/pavankrishnanarne/us-public-debt-quarterly-data-1996-present}{https://www.kaggle.com/datasets/pavankrishnanarne/us-public-debt-quarterly-data-1996-present}) and US Nominal GDP Quarterly Data(in Million \$)(\href{https://fred.stlouisfed.org/series/NGDPSAXDCUSQ}{https://fred.stlouisfed.org/series/NGDPSAXDCUSQ}), in quarters. Public Debt data was available from 1966-01-01 to 2023-01-01, and Nominal GDP data from 1966-01-01 to 2023-01-01.

  \bibliography{arXiv_bibliography.bib}

\end{document}